\documentclass[twocolumn,showpacs,preprintnumbers,superscriptaddress]{revtex4}
\usepackage{amsmath}
\usepackage{amsfonts}
\usepackage{amssymb}
\usepackage{graphicx}

\newcommand{\unity}{\ensuremath{{\rm 1\negthickspace l}{}}}

\begin{document}
\title{Time-Optimal Generation of Cluster States}%
\author{Robert Fisher}
\email{robert.fisher@ch.tum.de}
\affiliation{Department of Chemistry, Technische Universit\"at M\"unchen, Lichtenbergstrasse 4, D-85747 Garching, Germany}
\author{Haidong Yuan}
\affiliation{Department of Mechanical Engineering, MIT 3-160, Cambridge, Massachusetts 02139}
\author{Andreas Sp\"orl}
\affiliation{Department of Chemistry, Technische Universit\"at M\"unchen, Lichtenbergstrasse 4, D-85747 Garching, Germany}
\author{Steffen Glaser}
\affiliation{Department of Chemistry, Technische Universit\"at M\"unchen, Lichtenbergstrasse 4, D-85747 Garching, Germany}
\date{\today}
\begin{abstract}
The definition of a cluster state naturally suggests an implementation scheme: find a physical system with an Ising coupling topology identical to that of the target state, and evolve freely for a time of $\frac{1}{2J}$. Using the tools of optimal control theory, we address the question of whether or not this implementation is time-optimal. We present some examples where it is not and provide an explanation in terms of geodesics on the Bloch-sphere. 
\end{abstract}
\maketitle
\section{Introduction}
The one-way quantum computation model is an approach to quantum information processing where the evolution is driven by local operations and measurements only \cite{RaussendorfBriegel2,Nielsen}. In the experimental realization of such a model, preparation of the highly entangled initial state is therefore of primary concern. These initial states - so called cluster states - are also interesting in their own right, due to the favorable scaling of their entanglement properties \cite{DurBriegel,RaussendorfBriegel1}. To date, cluster states of four to six qubits have been realized experimentally in photonic systems \cite{Walther,Lu}, and are also actively pursued in other architechtures, e.g. ion traps \cite{Wunderlich}.

In this paper, the efficient generation of cluster states is studied using techniques from optimal control theory \cite{OC_Book1,OC_Book2}. Given an experimental framework, the aim of these techniques is to find the optimal set of controls to steer the system so that a desired target state or unitary gate is implemented. For systems consisting of two qubits, general analytical solutions exist for the construction of time-optimal unitary transformations \cite{Cartan,Haidong1} and state-to-state transfers \cite{Reiss1, Kramer} if fast local controls are available. For three or more qubits, analytical solutions are only known in some special cases. In \cite{threequbit1,Reiss2,Heitmann,Yuan_chain} it was shown that the time-optimal generation of indirect couplings and  trilinear Hamiltonians and the efficient transfer of order along Ising spin chains \cite{Haidong2,SpinChain1,SpinChain2} can be reduced to the problem of computing singular geodesics. In addition, powerful numerical methods \cite{JMRGrape} are available that make it possible to explore the physical limits of time-optimal control experiments in cases where no analytical solutions are known. The GRAPE algorithm \cite{JMRGrape} has been used for the generation of quantum gates \cite {QFT} and state-to-state transfers in NMR \cite{derJorginho}, as well as for superconducting qubits \cite{Frank_W}. Here, analytical and numerical techniques are used to investigate the problem of time-optimal cluster state preparation.

A cluster state is defined by a graph. To prepare the $n$-qubit cluster state corresponding to a graph $G$:
\begin{enumerate}
\item Prepare (locally) the initial state
\begin{align*}
|I_{n}\rangle := \left(\frac{|0\rangle+|1\rangle}{\sqrt{2}}\right)^{\otimes n}.
\end{align*}
\item Evolve under the Ising Hamiltonian
\begin{align}
\label{IsingHamiltonian}
H_{d} = \frac{\pi J(t)}{2} \sum_{\langle a,a' \rangle} \left(\unity+\sigma_{z}^{(a)}\right)\left(\unity-\sigma_{z}^{(a')}\right)
\end{align}
for a time such that $\int_{0}^{T}J(t)\,dt=\frac{1}{2}$. The sum is over all edges of $G$, where each edge connects the qubit pair $\langle a,a' \rangle$.
\end{enumerate}
We consider $n$-qubit systems of the form
\begin{align*}
H=H_d+\sum_{j}u_{j}\,H_{c}^{(j)},
\end{align*}
where the $u_{j}$ are time-dependent functions to be chosen, and the $H_{c}^{(j)}$ characterize the available controls. Two different control settings will be considered:
\begin{enumerate}
\item[(i)] Local x and y control on each qubit:
\begin{align}
\label{localControls}
H_{c}^{(2j-1)}=\frac{1}{2}\,\sigma_x^{(j)}, \:\: H_{c}^{(2j)}=\frac{1}{2}\,\sigma_y^{(j)}
\end{align}
where $j\in\{1,2,...,2n\}$.
\item[(ii)] A single global x control:
\begin{align}
\label{globalControl}
H^{(1)}_{c}=F_{x}:=\frac{1}{2}\sum_{j=1}^{n}\sigma_{x}^{(j)}.
\end{align}
\end{enumerate}
In the following, we allow for fast local controls, i.e. the functions $u_{j}$ are unrestricted.
\section{3 Completely-Coupled Qubits}
\label{section2}
\begin{figure}[!htbp]
\includegraphics[width=0.14\columnwidth]{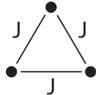}
\caption{$K_{3}$ coupling graph.}
\label{k3CouplingGraph}
\end{figure}

To shed some light on when a speedup may be possible, we begin with a symmetric 3-qubit system which will prove analytically tractable. The qubits are Ising-coupled according to the complete coupling graph $K_{3}$ \cite{GraphTheory}, illustrated in fig. \ref{k3CouplingGraph}.
Without loss of generality we will drop the local terms in (\ref{IsingHamiltonian}), as only the entangling part of the operation contributes to the time required. Furthermore $J$ is assumed to be constant. The $K_{3}$ cluster state is then defined as
\begin{align*}
\left|T_{3}\right\rangle:=\exp\left(-i\frac{1}{2J}H_{d}\right)\left|I_{3}\right\rangle,
\end{align*}
where
\begin{align*}
H_d=\frac{\pi J}{2}\left(\sigma_{z}^{(1)}\sigma_{z}^{(2)}+\sigma_{z}^{(2)}\sigma_{z}^{(3)}+\sigma_{z}^{(1)}\sigma_{z}^{(3)}\right).
\end{align*}
The preparation of this state poses the following control problem: maximize
\begin{align}
\label{fidelity}
F(U):=\left|\langle T_{3}|\cdot\left(U|I_{3}\rangle\right)\right|
\end{align}
subject to equation of motion $\dot{U}=-iHU$. In the first instance we will allow for full local control on the qubits and thus specify $H_c$ according to (\ref{localControls}).

To maximize the fidelity defined in (\ref{fidelity}) in the shortest possible time, we first apply a numerical gradient-ascent algorithm, as detailed in \cite{JMRGrape}. Over a comprehensive range of initial conditions the controls are updated iteratively, incrementing $F$ to a local maximum.
\begin{figure}[!bp]
\hspace{-0.1cm}\includegraphics[width=0.95\columnwidth]{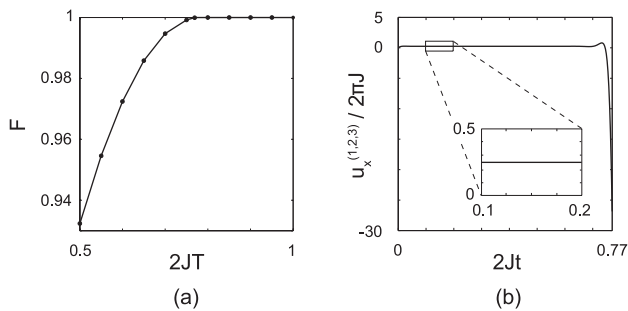}
\caption{(a) Maximum achievable fidelity as a function of the transfer time. (b) A sample solution at the minimal time for initial controls $u_{x}^{j}(t)=u_{y}^{j}(t)=0$. The y controls are omitted as they remain zero, and the x controls are identical on each qubit due to the permutation symmetry of the drift Hamiltonian and the initial and target states.}
\label{grape3q}
\end{figure}
The minimal time found by the algorithm to generate $|T_{3}\rangle$ is approximately $0.77 \times \frac{1}{2J}$, as seen in fig. \ref{grape3q}a. The significance of this particular value will become clearer later. Fig. \ref{grape3q}b shows a numerically optimised solution close to the minimal time. We find in all observed cases that the maximum fidelity does not depend on whether the controls are specified according to (\ref{localControls}) or (\ref{globalControl}).

Motivated by these symmetric solutions, we now restrict ourselves to control setting (ii), specifying $H_c$ according to (\ref{globalControl}). This control Hamiltonian, in addition to the drift Hamiltonian and the inital and target states, is symmetric under the cyclic permutation operator 
\begin{align*}
S:=\left[\begin{array}{cccccccc}
1 & 0 & 0 & 0 & 0 & 0 & 0 & 0 \\
0 & 0 & 1 & 0 & 0 & 0 & 0 & 0 \\
0 & 0 & 0 & 0 & 1 & 0 & 0 & 0 \\
0 & 0 & 0 & 0 & 0 & 0 & 1 & 0 \\
0 & 1 & 0 & 0 & 0 & 0 & 0 & 0 \\
0 & 0 & 0 & 1 & 0 & 0 & 0 & 0 \\
0 & 0 & 0 & 0 & 0 & 1 & 0 & 0 \\
0 & 0 & 0 & 0 & 0 & 0 & 0 & 1
\end{array}\right]
\end{align*}
and the persymmetry operator
\begin{align*}
P:=\left(\sigma_{x}\right)^{\otimes 3},
\end{align*}
which themselves commute. The dynamics are thus restricted to the simultaneous eigenspace of $S$ and $P$ corresponding to the eigenvalue pair $\{1,1\}$. Transforming to a new basis composed of the simultaneous eigenstates of $S$ and $P$ makes this explicit  \cite{BanwellPrimas}. The Hamiltonians are now diagonalized into $2\times 2$ blocks. As we need only consider the $\{1,1\}$ block, the state transfer problem can be reduced to
\begin{align}
\label{2dtransfer}
\left|I_3'\right\rangle
=\frac{1}{2}
\left[\!\begin{array}{r}
\sqrt{3} \\ 1
\end{array}\right]
\:\longrightarrow\:
\left|T_3'\right\rangle
=\frac{1}{2}
\left[\!\begin{array}{r}
\sqrt{3} \\ -1
\end{array}
\right],
\end{align}
under the Hamiltonian $H'=H'_{d}+u(t)H'_{c}$, where
\begin{align}
\label{2dHdHc}
H'_{d}=\frac{\pi J}{2}
\left[\!\begin{array}{r r}
-1 & 0 \\ 0 & 3 
\end{array}\right], \:\:\:\:\:
H'_{c}=\frac{1}{2}
\left[\!\begin{array}{r r}
2 & \sqrt{3} \\ \sqrt{3} & 0 
\end{array}\right].
\end{align}

\begin{figure}[!htbp]
\includegraphics[width=0.8\columnwidth]{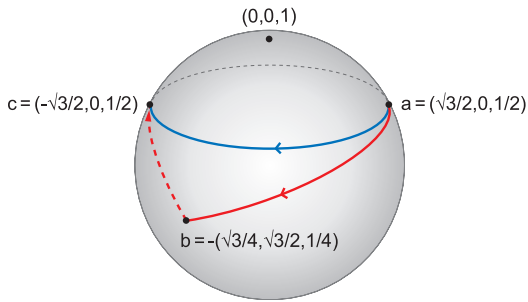}
\caption{(Color online) Transfer of eqn. (\ref{2dtransfer}) on the Bloch sphere. The $u=0$ solution (blue line) transfers $a$ to $c$ in a time of $\frac{1}{2J}$. A time-optimal solution (red line) transfers $a$ to $b$ in $\frac{2}{3\sqrt{3}J}$, followed by a hard pulse (dashed red line) from $b$ to $c$ in negligible time.}
\label{BlochSphere1}
\end{figure}

\begin{figure}[!bp]
\includegraphics[width=0.9\columnwidth]{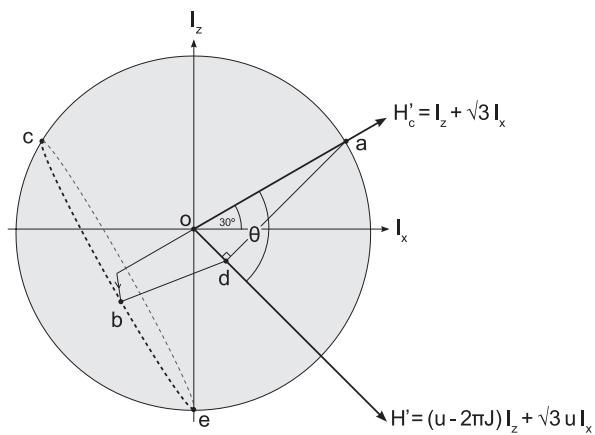}
\caption{Geometric constructions used. All points shown here lie in the $I_{x}$-$I_{z}$ plane \textit{except} for $b$, which lies above it on the upper surface of the sphere.}
\label{fig4}
\end{figure}

In order to consider the problem geometrically, we represent the transfer on the Bloch sphere by projecting onto the axes $I_{j}:=\sigma_j/2$, as illustrated in fig. \ref{BlochSphere1}.
The initial and final states $|I_3'\rangle$ and $|T_3'\rangle$ are identified with vectors $a=\frac{1}{2}(\sqrt{3},0,1)$ and $b=\frac{1}{2}(-\sqrt{3},0,1)$, respectively, while the Hamiltonians $H'_{d}$ and $H'_{c}$ correspond to (unnormalized) rotation axes $(0,0,-2\pi J)$ and $(\sqrt{3},0,1)$, respectively. The trivial solution is to set $u(t)=0$ and rotate about $H'_{d}$ for $\frac{1}{2J}$ units of time, but faster solutions may exist. Motivated by the numerical results in Fig. \ref{grape3q}b, we first restrict $u(t)$ to the following form:
\begin{enumerate}
\item Constant pulse $u$ over time interval $[0,T]$.
\item Hard pulse of angle $\phi$ at time $T$.
\end{enumerate}
After deriving a time-optimal solution in this setting, we will show that it remains time-optimal when the restrictions are removed and general time-varying pulses are considered. The solution is
\begin{align}
\label{timeOptSolution}
(u,\phi,T)=\left(\frac{\pi J}{2},\frac{-\pi}{4},\frac{2}{3\sqrt{3}J}\right),
\end{align}
which explains the value of $T\approx 0.77 \times \frac{1}{2J}$ obtained numerically.
To demonstrate the time-optimality of this solution, we can consider the geometric constructions in fig. \ref{fig4}.
Starting from $a$, the task is to transfer the state to any point $b$ on the circle $\widehat{bce}$ obtained by rotating about $H'_{c}$. The hard pulse then transfers $b$ to $c$ in an arbitrarily small time. The choice of constant $u$ specifies a rotation axis $H'$, at an angle $\theta$ to $H'_c$. The transfer time to be minimized is
\begin{align*}
T=\frac{\angle adb}{|H'|},
\end{align*}
where $\angle adb$ is the angle swept out by the Bloch vector and the length $|H'|$ gives its angular velocity of rotation. Using simple geometry these quantities are expressed in terms of $\theta$ as
\begin{align*}
\angle adb=2\arcsin\left(\frac{\sqrt{3}}{2\sin\theta}\right),\:\:\: |H'|=\frac{\sqrt{3}\pi J}{\sin\theta}
\end{align*}
so that the time is
\begin{align*}
T=\frac{2}{\sqrt{3}\pi J}\sin\theta\arcsin\left(\frac{\sqrt{3}}{2\sin\theta}\right).
\end{align*}
Noting that $\theta$ must lie in the interval $[\frac{\pi}{3},\frac{2\pi}{3}]$ for intersection with the circle, we find the maximum at $\theta=\frac{\pi}{2}$. The time required is $T_{\text{min}}=\frac{2}{3\sqrt{3}J}$. The angle of the final hard pulse is $-\frac{\pi}{2}$ in this picture, but since $|H'_{c}|=2$ the angle around a normed axis is $\phi=-\frac{\pi}{4}$.

\begin{figure}[!pb]
\hspace{0.1cm}\includegraphics[width=0.95\columnwidth]{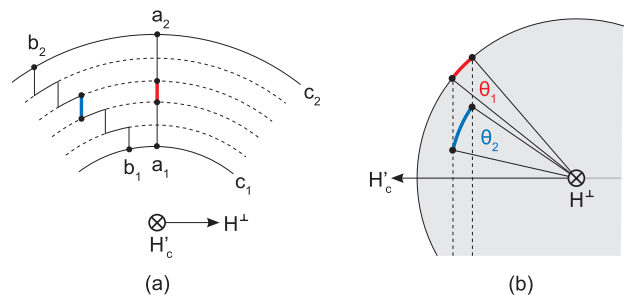}
\caption{(Color online) (a) An arbitrary path can be decomposed into rotation about $H'_c$ and rotation about the orthogonal axis $H^{\bot}$ via the Trotter decomposition. (b) Rotating our viewpoint by $90^{\circ}$, we see that the optimal trajectory from $a_{1}\rightarrow a_{2}$ minimizes the angle rotated through in each segment (red), when compared to a generic trajectory (blue), ie. $\theta_{1}<\theta_{2}$.}
\label{fig5}
\end{figure}

It remains to show that $\widehat{bce}$ cannot be reached in less than $T_{\text{min}}$ when allowing for time-varying pulses. For this we introduce
\begin{align*}
H^{\bot}:=\frac{\sqrt{3}\pi J}{2}\left(-\sqrt{3}I_z+I_x\right),
\end{align*}
which is simply the rotation axis orthogonal to $H'_{c}$. We consider a generic time-varying control $u(t)$. Our aim is to compare each segment of this generic path to the optimal one. Let $c_1$ and $c_2$ be two circles generated by rotating about $H'_{c}$, chosen to be close enough so that $u(t)$ is well approximated by a constant in the interval between them. We consider the time required to travel from  $c_1$ to $c_2$ along two different paths: our proposed optimal solution $a_{1} \rightarrow a_{2}$, obtained by rotating purely about $H^{\bot}$, and a generic path $b_{1} \rightarrow b_{2}$. Suppose the evolution along $b_{1} \rightarrow b_{2}$ takes a time $\Delta$. This evolution can be decomposed according to the Trotter formula
\begin{align}
\label{trotter}
e^{-i\Delta(H^{\bot}+vH'_{c})}=\lim_{n\rightarrow\infty}\left(e^{-i\frac{\Delta}{n}H^{\bot}}e^{-i\frac{v\Delta}{n}H'_{c}}\right)^n,
\end{align}
which is represented graphically in fig. \ref{fig5}a.
Note that the time required for the operation on the right-hand side of (\ref{trotter}) is still $\Delta$, as the evolutions along $H'_{c}$ can be arbitrarily fast. The time $\frac{\Delta}{n}$ in a single segment is equal to the angle swept out divided by the norm of $H^{\bot}$. This angle is minimized in every segment when travelling from $a_{1} \rightarrow a_{2}$, as fig. \ref{fig5}b illustrates. The time-optimal solution is therefore to rotate purely about $H^{\bot}$, which corresponds exactly to solution (\ref{timeOptSolution}). The speedup here arises from the fact that $H'_{d}$ and $H'_{c}$ are not orthogonal, which can be carried over into higher dimensions.
\section{Higher Dimensional Cases}
\begin{figure}[!htbp]
\includegraphics[width=0.7\columnwidth]{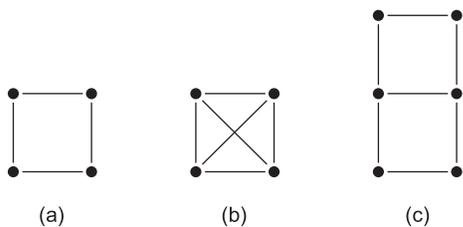}
\caption{Some of the graphs considered: (a) $C_4$, (b) $K_4$, (c) $G_{2,3}$.}
\label{fig6}
\end{figure}

\begin{figure}[!htbp]
\includegraphics[width=0.95\columnwidth]{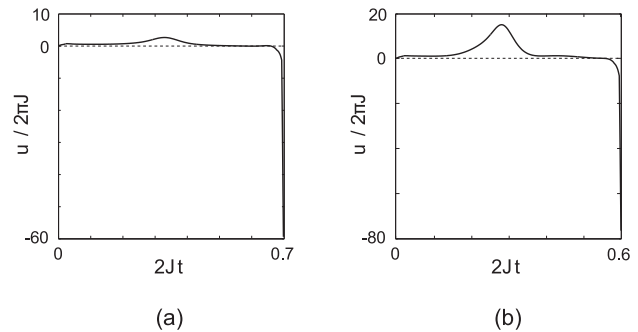}
\caption{Numerically optimised sequences for (a) $K_5$ and (b) $K_7$.}
\label{grapeK7s}
\end{figure}

Finally we provide the minimal times for cluster state preparation on a variety of graphs, some of which are illustrated in fig. \ref{fig6} to clarify our notation.
The times are evaluated numerically using the GRAPE algorithm, and are included in Table \ref{resultsTable}. 
\begin{table}[bp]
\centering
\begin{tabular}{c c c c}
\hline
$\:\:$ Graph $\:\:$ & $\:\:$ $d$ $\:\:$ & $\:\:$ $|\langle \hat{H}'_d | \hat{H}'_c \rangle|$ $\:\:$ & $\:\:$ $T_{\text{min}}$ ($\frac{1}{2J}$) $\:\:$ \\
\hline \hline
$K_3$     & 2  & 0.2  & 0.77 \\
$L_3$     & 3  & 0    & 1.00 \\
$K_4$     & 3  & 0    & 0.91 \\
$C_4$     & 4  & 0    & 1.00 \\
$K_5$     & 3  & 0.1  & 0.70 \\
$K_6$     & 4  & 0    & 1.00 \\
$G_{2,3}$ & 14 & 0    & 1.00 \\
$K_7$     & 10 & 0.07 & 0.60 \\
\hline
\end{tabular}
\caption{Minimal times calculated by the GRAPE algorithm, with an estimated numerical accuracy of $\pm 1$ on the last digit. As for the 3-qubit example under control setting (ii), included in the first row as a reference, the drift and control can be reduced to matrices $H'_d$ and $H'_c$ of size $d\times d$. $\hat{H'}$ indicates a rescaling to unit norm.}
\label{resultsTable}
\end{table}
For all of the 4-qubit graphs, both control schemes (i) and (ii) were considered, yielding exactly the same minimal times in each case. For the larger graphs only control scheme (ii) was considered, allowing us to reduce the problem to a dimension $d$ using a symmetry-adapted basis, as in Sec. \ref{section2}. The times listed here hold not just for the target state but its entire local unitary orbit, which may include other entangled states of interest \cite{EisertBriegel}. While a Bloch sphere analysis is not possible for $d>2$, we find that a connection persists between the non-orthogonality of $H'_{d}$ and $H'_{c}$ and the minimal time - in all cases considered, a speedup is possible when $H'_{d}$ and $H'_{c}$ are non-orthogonal. Minimal time solutions found by the algorithm for $K_5$ and $K_7$ are shown in fig. \ref{grapeK7s}.
We observe that the effective hard rotation pulse at time $T$ is still present, while the preceding pulse shape is no longer constant.
\section{Summary}
For several known two-qubit quantum gates, the `do nothing' operation, i.e. the evolution under a given coupling Hamiltonian, is time-optimal. For example, this is the case for a SWAP gate in the presence of an isotropic Heisenberg coupling \cite{SWAP}. Hence it may be surprising to find that it is possible to create cluster states faster than the time required by the straightforward implementation their definition implies. Here we provide examples where this is the case. In particular for three Ising-coupled qubits with identical coupling constants, the problem of time-optimal cluster state generation could be solved analytically via geodesics on a sphere. The numerical techniques used here can also be applied to the problem of cluster state generation in an arbitrary coupling topology, where real experimental values for the coupling constants can be used. This will be addressed in a future work.
\section*{Acknowledgements}
This work has been supported by the Elite Network of Bavaria programme QCCC, the integrated EU project QAP, and the Deutsche Forschungsgemeinschaft incentive SFB-631. R.F. and S.G. would like to thank Christof Wunderlich and Thomas Schulte-Herbr\"uggen for valuable discussions.
\end{document}